\documentclass[useAMS,usenatbib]{mn2e}

\usepackage{times}
\usepackage[latin1]{inputenc}
\usepackage{amssymb}
\usepackage{amsfonts}
\usepackage{amsmath}
\usepackage{mathrsfs}
\usepackage{txfonts}
\usepackage{graphicx}
\usepackage{longtable}

\newcommand{\msolyr}{\ensuremath{M_\odot \, {\rm yr}^{-1}}}

\newcommand{\mdotw}{\ensuremath{\dot{M}_{\rm w}}}

\newcommand{\ltapprox}{\raisebox{-0.5ex}{$\,\stackrel{<}{\scriptstyle\sim}\,$}}
\newcommand{\gtapprox}{\raisebox{-0.5ex}{$\,\stackrel{>}{\scriptstyle\sim}\,$}}

\begin{document}

\title[Radio and X-ray emission from disc winds]{Radio and X-ray emission from disc winds in radio-quiet quasars}

\author[K.C. Steenbrugge, E.J.D. Jolley, Z. Kuncic, and K.M. Blundell]{
K. C. Steenbrugge$^{1,2}$\thanks{E-mail: katrien.steenbrugge@gmail.com}, 
E.J.D. Jolley$^{3}$, Z. Kuncic$^{3}$ and K.M. Blundell $^{2}$ \\
$^{1}$Instituto de Astronom\'ia, Universidad Cat\'olica del Norte, Avenida Angamos 0610, Casilla 1280, Antofagasta, Chile\\
$^{2}$Department of Physics, University of Oxford, Keble Road, Oxford OX1 3RH, UK\\
$^{3}$School of Physics, University of Sydney, Sydney NSW 2006, Australia\\
}
\date{}

\pagerange{\pageref{firstpage}--\pageref{lastpage}} \pubyear{2008}

\maketitle

\label{firstpage}

\begin{abstract}
It has been proposed that the radio spectra of radio-quiet quasars is
produced by free-free emission in the optically thin part of an
accretion disc wind. An important observational constraint on this
model is the observed X-ray luminosity. We investigate this constraint
using a sample of PG radio-quiet quasars for which XMM-{\it Newton}
EPIC spectra are available. Comparing the predicted and measured
luminosities for 0.5, 2 and 5~keV, we conclude that all of the studied
PG~quasars require a large hydrogen column density absorber, requiring
these quasars to be close to or Compton-thick. Such a large column
density can be directly excluded for PG~0050+124, for which a
high-resolution RGS spectrum exists. Further constraint on the column
density for a further 19 out of the 21 studied PG~quasars comes from
the EPIC spectrum characteristics such as hard X-ray power-law photon
index and the equivalent width of the Fe~K$\alpha$ line; and the small
equivalent width of the C~IV absorber present in UV spectra. For 2
sources: PG~1001+054 and PG~1411+442 we cannot exclude that they are
indeed Compton-thick, and the radio and X-ray luminosity are due to a
wind originating close to the super-massive black hole. We conclude
that for 20 out of 22 PG~quasars studied free-free emission from
a wind emanating from the accretion disc cannot mutually explain the
observed radio and X-ray luminosity.
\end{abstract}


\section{Introduction}

Currently, the origin of the nuclear radio emission from radio-quiet
quasars (RQQs), which consists of 90$\%$ of the optically detected
quasars \citep{ivezic02}, is not understood. The luminosity of the
nuclear radio emission and its compactness is similar to the nuclear
radio luminosity, brightness temperature and compactness of radio-loud
quasars \citep{blundell98,ulvestad05}. Likewise the variability
\citep{barvainis05} is similar between both classes. For radio-loud
quasar the nuclear radio emission is explained by the superposition of
different jet-components \citep{cotton80}. However, radio-quiet
quasars lack radio emission on large scales which are indicative of
expanding lobes fed by jets. A possibility is that radio-quiet quasars
do have much weaker jets, which do not escape the inner 1~kpc of the
host galaxy \citep{miller93,kuncic99}. However most RQQs studied with
VBLI show unresolved radio emission, even at mas scales, indicating an
emission volume of a few pc$^3$
\citep{blundell98,ulvestad05}. Therefore, the suggestion that the
radio emission from nuclei in RQQ is due to a superposition of jet
knots, seems unrealistic for those RQQ where there is no evidence for
a jet.

Recently, two alternative models explaining the radio emission in RQQs
have been proposed. \cite{Blundell07} proposed that radio emission in
radio-quiet quasars originates in a optically thin plasma with a
temperature of order 10$^7$~K, such that the plasma is completely
ionised. This plasma has a density high enough that free-free
processes dominate. Such a plasma can explain the observed brightness
temperature of the nuclei of RQQs, if the optical depth is close to
unity. This model explains the flat radio spectra observed in
quasars. The authors assume that the wind is launched from the
accretion disc at a radius $\approx 10^{-3}$ pc, and becomes optically
thin at a photospheric radius of 0.1$-$1~pc. In this model the
mass-loss rates in these winds are significant, and these winds could
have an effect on the feedback processes operating in the host galaxy.

An alternative model was proposed by \cite{LaorBehar08}, who model the
radio and X-ray emission as due to coronal emission, similar to
coronal emission observed in stars. They base this model on the
observation that there is a rather tight correlation between the radio
and X-ray luminosity: $L_{\rm R}$ $\sim$ 10$^{-5}$$L_{\rm X}$, similar
to the G\"{u}del-Benz relation for coronally active stars. For
coronally active stars it has been shown that this relation is due to
magnetic heating of the corona. \cite{LaorBehar08} therefore propose
that the radio and X-ray emission from the nuclei of RQQs is due to
magnetic heating of the corona, presumed to be located above the
accretion disc. As with the optically thin wind model, the predicted
spectrum at a few GHz is flat.

In this paper we test the optically thin wind model as proposed by
\cite{Blundell07}. For an optically thin wind one can predict the
X-ray luminosity for a given radio luminosity. One can then compare
the predicted X-ray luminosity to the measured X-ray luminosity. The
wind model overpredicts the soft X-ray luminosity \citep{LaorBehar08},
unless intervening absorbing material is present in the outflow. The
absorption could arise from a radiatively cooled part of the wind
further from the accretion disc. The X-ray continuum is generally well
fit by a power-law and a (modified) black body at soft energies. The
black body fits the soft excess emission probably from the accretion
disc or due to ionised reflection. The power-law component is
generally assumed to be Comptonised emission from the corona located
somewhere close to the accretion disc. Alternatively, a good fit can
be obtained with a broken power-law with a break energy at
$\sim$1.5~keV.

X-ray absorbing outflows (see \citealt{Crenshaw03} for a review), are
especially well studied in Seyfert 1 galaxies, due to their X-ray
brightness, which allows for high-resolution X-ray
spectroscopy. Generally the absorption consists of 2 or more
ionisation components, spanning a range in ionisation parameter of 3
or more orders of magnitude, and have outflow velocities similar to
those measured at UV wavelengths of $v_{\rm w}$$\sim$ 100 $-$
1000~km~s$^{-1}$ \citep{Kaastra00,Kaspi01,Kaastra02}. This outflowing
gas absorbs the continuum and produces line and edge absorption,
allowing for their kinematics to be studied.  Warm absorbers have been
detected in more than $50$\% of Seyfert 1 galaxies \citep{Crenshaw03,
  Reynolds97, George98}. A similar fraction of RQQs might have a warm
absorber (\citealt{porquet04,Piconcelli05}, but see also
\citealt{Brocksopp06}).
 
The exact location of the absorbing gas, even in the well studied
Seyfert~1 galaxies, as well as its origin, are still not well
constrained \citep{steenbrugge09,gabel05}, and two origins have been
proposed. The absorbing outflows are thought to originate either from
the accretion disc (e.g. \citealt{MurrayChiang95,
  Murray95,Elvis00,Proga00,everett07}), or from the molecular torus
(e.g. \citealt{KrolikKriss01,Ashton04,Blustin05}). Similarly, the
ionisation structure of the absorber is still poorly constrained, and
can currently be well fit with several discrete ionisation components,
modelling photoionised clumps (e.g. \citealt{Arav05, KrolikKriss01,
  Rozanska06}) or a continuous ionisation structure
(e.g. \citealt{Steenbrugge05, Behar09}). The total X-ray observed
hydrogen column density in Seyfert~1 galaxies spans at least 2 orders
of magnitude, from 4.9$\times$10$^{24}$ m$^{-2}$ in Mrk~279
\citep{ebrero10} to 3.8$\times$10$^{26}$ m$^{-2}$ in NGC~3783
\citep{CheloucheNetzer05}.

Approximately 22\% of optically detected quasars exhibit large
hydrogen absorbing column densities, the broad absorption line (BAL)
quasars \citep{HewettFoltz03}. The total column density derived from
UV observations is $\sim$5$\times$10$^{27}$ m$^{-2}$ \citep{krolik99},
while \cite{blustin08} find an average X-ray total column density of
2$\times$10$^{28}$ m$^{-2}$. It is currently unknown whether the
detection of these broad absorption lines is due to orientation
\citep{weymann91}, or whether BALQSOs are an evolutionary phase of
QSOs \citep{briggs84}.

This paper is organised as follows. Section \ref{bremss} summarises
the bremsstrahlung disc wind model for radio emission and the
associated X-ray luminosity predicted by this model. We also show that
this wind can radiatively cool and form an absorber. In the following
section we describe the sample we use and compare the predicted and
measured X-ray luminosities. In Section \ref{abs} we model the
required absorption to explain the difference between the predicted
and measured X-ray luminosity. In Section~\ref{discussion} we show
that absorption cannot be the origin of the difference between
predicted and measured X-ray luminosity, with the exception of
PG~1001+054 and PG~1411+442. Our conclusions are given in Section
\ref{conclusion}.

\section{Disc Wind Emission and absorption}\label{bremss}
\subsection{Emission model}

\begin{center}
\begin{figure}
\includegraphics[width=8truecm]{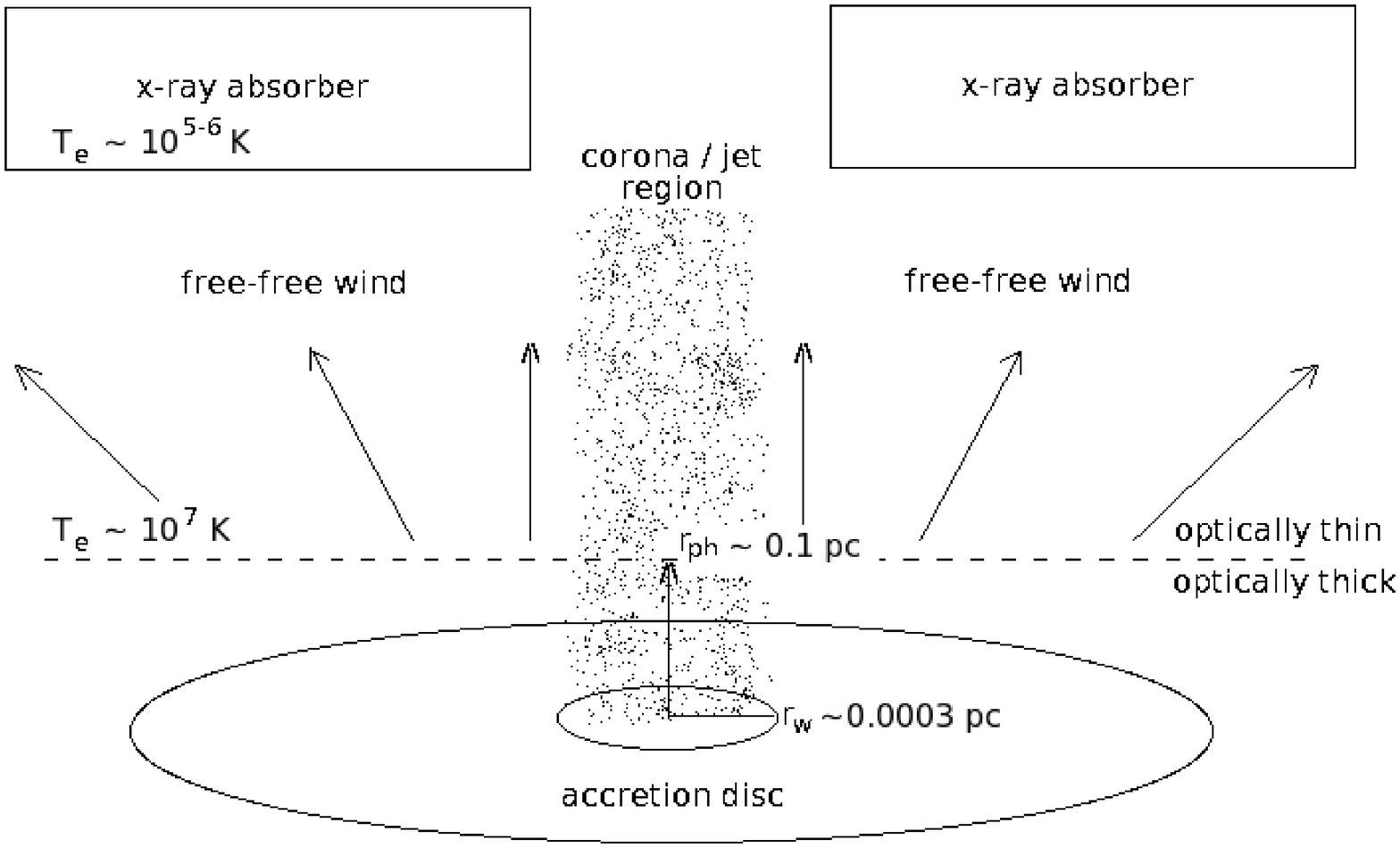}
\caption{Schematic illustration of the disc wind and X-ray absorber
  (not to scale). $r_{\rm w}$ is the characteristic radius at which
  the free-free wind is launched from the disc, $r_{\rm ph}$ is the
  distance at which the wind becomes transparent to free-free
  absorption, and $T_{\rm e}$ is the electron temperature.}
\label{fig:fig1}
\end{figure}
\end{center}

Following \citet{Blundell07}, we consider a thermal accretion disc
wind at radii $r \gtapprox r_{\rm w}$, where $r_{\rm w}$ is the
launching radius of the wind.  A hot disc wind that is optically-thin
to both electron scattering and free-free absorption can exist beyond
a photospheric radius $r_{\rm ph} \simeq 0.1 - 1 \,{\rm pc}$. In this
part of the wind, with mass outflow rate $\mdotw$, the specific
bremsstrahlung luminosity is
\begin{equation}\label{Lnu}
L_\nu = 7.8\times 10^9 \bar g_{\rm ff} f_\Omega^{-1} r_{\rm ph}^{-1} T_{\rm e}^{-1/2} \exp(-h\nu / kT_{\rm e}) \mdotw^2 v_{\rm w}^{-2}
\, \, \, \rm erg\,s^{-1}
\end{equation}
where $f_\Omega = \Omega /4\pi \ltapprox 1$ is the geometrical
covering factor of the outflow, $v_{\rm w}$ is the outflow velocity at
radius $r_{\rm w}$ and $\bar g_{\rm ff}$ is the velocity averaged
free-free Gaunt factor \citep{RL}. At radio frequencies, $\bar g_{\rm
  ff} \simeq 10$ and at X-ray energies, $\bar g_{\rm ff} \simeq 1$. A
schematic illustration of the model is presented in
Fig.~\ref{fig:fig1}.

Assuming the radio and X-ray emission are produced by the same
mechanism from the same source, eq. 1 also applies at X-ray
energies. The expected X-ray luminosity is thus:
\begin{equation}
L_\nu \approx \frac{L_\nu}{10} \exp \left(-\frac{h\nu_X}{k T_e}\right)
\end{equation}
where the factor $1/10$ is the ratio of the Gaunt factors at these
energies, and we ignore the exponential factor at the radio frequency,
as it is negligible.  We will use this formula to predict the X-ray
luminosity at 0.5, 2 and 5~keV.

\subsection{Absorption model}
Considering that the X-ray luminosity is over-predicted in this model
\citep{LaorBehar08}, we here describe a model of how the hot optically
thin wind can cool to produce an absorber and derive that the required
density is within a reasonable range.

Outflowing diffuse gas can radiatively cool to produce the absorber
observed in Seyfert~1 galaxies and RQQs provided that the radiative
cooling timescale $t_{\rm c}$ is much less than the dynamical cooling
timescale $t_{\rm dyn}$:
\begin{equation}\label{timescales}
t_{\rm c} \ll t_{\rm dyn} = \frac{R}{v_{\rm w}}
\end{equation}
The bremsstrahlung cooling time can be expressed as 
\begin{equation}\label{tb}
t_{\rm c} = 1.0 \times 10^{-12} f_\Omega r_{\rm w}^2 \bar g_{\rm b}^{-1} T_{\rm e}^{1/2} v_{\rm w} / \mdotw  \,\,\,\rm s
\end{equation}
where $\bar g_{\rm b} \simeq 1.2$ is the frequency-averaged Gaunt
factor, and $T_{\rm e} = 10^7$ K is the temperature of the emitting
plasma near the wind photosphere. For canonical parameters, we use the
values $v_{\rm w} = 500$ km s$^{-1}$, $r_{\rm w} = 10^{15}$ cm,
$f_\Omega = 0.1$, and $R = r_{\rm ph} = 0.1$ pc
\citep{Blundell07}. Eqs. \ref{timescales} and \ref{tb} imply that the
wind mass-loss rate must satisfy
\begin{equation}\label{mdotwmin}
\mdotw \gg 4 \times 10^{-10} \,\,\,\msolyr
\end{equation} 
in order for the gas to radiatively cool on a timescale shorter than
the dynamical timescale.

At the base of the disc wind, the electron number density $n_{\rm e}$
can be related to the mass outflow rate using the continuity equation,
\begin{equation}
 \mdotw = 2\pi f_\Omega r_{\rm w}^2 v_{\rm w} m_{\rm p} n_{\rm e} \simeq 8 \times 10^{-13} n_{\rm e} \, \, \, \msolyr
\end{equation}
which, even for a rarefied wind with $n_{\rm e} \simeq 10^5 \, \rm
cm^{-3}$, clearly satisfies Eq. \ref{mdotwmin}. Hence, the emitting
outflow from the disc wind can cool and form a partially ionised
absorber as observed in a few RQQs and about half of Seyfert~1
galaxies.

\section{Data}

A useful sample of optically-selected radio-quiet quasars is the
Palomar-Green (PG) Bright Quasar Survey \citep{SchmidtGreen83}. Our
sample of PG~quasars is dictated by the need of good medium-resolution
EPIC XMM-{\it Newton} data or high signal-to-noise high-resolution
X-ray spectra. We therefore chose to use the RQQs in the sample of
PG~quasars which were studied in detail by \cite{Brocksopp06} and
added PG~0050+124 which has a high-signal-to noise XMM-{\it Newton}
RGS spectrum \citep{Costantini08}, PG~0844+349 and
PG~2214+139. PG~0844+349 was studied in detail by \cite{brinkmann06}
and \cite{gallo10} and PG~2214+139 by \cite{Piconcelli04}. This gives
us a sample of 22 RQQs. For this subset of PG~quasars we utilise the
5~GHz radio luminosities compiled by \citet{LaorBehar08} which are
based on observations with the Very Large Array by
\citet{Kellermann89} and \citet{Kellermann94}. The studied PG~quasars
all have a redshift $z < 0.5$ \citep{BorosonGreen92}. In this sample,
$3$ have measured UV C\,IV equivalent line widths greater than $1$
\AA~\citep{LaorBehar08}, which suggests the soft X-ray band is likely
to be affected by X-ray absorption. These are : PG~1001+054,
PG~114+445 and PG~1411+442. The final quasar sample is given in
Table~\ref{tab:tab1}, which lists the redshift, the 5~GHz radio and
the 2$-$5 and 0.3$-$10~keV X-ray luminosity.

\begin{center}
\begin{table}
\caption{Properties of the PG~quasars studied in this paper. The
  redshifts and radio $\nu$L$_{\nu}$ data is taken from Laor \& Behar
  (2008), the X-ray $\nu$L$_{\nu}$ for the 2$-$5 and 0.3$-$10~keV band
  is taken from Brocksopp et al. (2006). For PG~0050+124, we use the
  unabsorbed 0.5$-$2~keV luminosity determined from the RGS spectra
  (Costantini et al. 2008). For PG~0844+349 we use the 2$-$10 and
  0.5$-$2~keV luminosities for the 2001 high flux state as given by
  Gallo et al. (2010), and for PG~2214+139 we list the 0.5$-$2 and
  2$-$10~keV luminosity given by Piconcelli et al. (2004).}
\label{tab:tab1}
\begin{tabular}{lccccccccccccc}\hline\hline 
name        & redshift & $\nu$L$_R^{a}$  & $\nu$L$_X^{b}$ & $\nu$L$_X^{b}$ \\
            &          & 5~GHz       & 2$-$5 keV  & 0.3$-$10 keV \\\hline
PG~0050+124 & 0.0587   & 9.33        & $-$        & 0.7$^{c}$   \\
PG~0844+349 & 0.0644   & 0.65        & 0.7$^{d}$  & 1$^{c}$     \\
PG~0947+396 & 0.2059   & 11.7        & 1.18       & 5.4         \\
PG~0953+414 & 0.2341   & 95.5        & 2.89       & 15.7        \\
PG~1001+054 & 0.1610   & 24.0        & 0.025      & 0.1         \\
PG~1048+342 & 0.1667   & $<$4.07     & 0.55       & 2           \\
PG~1114+445 & 0.1438   & 4.47        & 0.57       & $-$$^{e}$   \\
PG~1115+407 & 0.1542   & 4.47        & 0.5        & 3.3         \\
PG~1116+215 & 0.1765   & 158.5       & 1.79       & 10.3        \\
PG~1202+281 & 0.1654   & 5.49        & 1.36       & 5.5         \\
PG~1216+069 & 0.3318   & 416.9       & 2.14       & 2.3         \\
PG~1322+659 & 0.1676   & 6.31        & 0.64       & 4.3         \\
PG~1352+183 & 0.1508   & 5.75        & 0.66       & 3.5         \\
PG~1402+261 & 0.1643   & 18.6        & 0.82       & 5.4         \\
PG~1411+442 & 0.0897   & 3.16        & 0.02       & 1.2         \\
PG~1415+451 & 0.1133   & 3.16        & 0.21       & 1.2         \\
PG~1427+480 & 0.2203   & $<$7.59     & 0.89       & 4.3         \\
PG~1440+356 & 0.0777   & 8.71        & 0.26       & 2.2         \\
PG~1444+407 & 0.2676   & $<$9.77     & 0.76       & 5.2         \\
PG~1543+489 & 0.4009   & 27.5        & 0.71       & 4.6         \\
PG~1626+554 & 0.1317   & 3.31        & 0.85       & 3.6         \\ 
PG~2214+139 & 0.0657   & 1.07        & 0.48$^{d}$ & 0.39$^{c}$   \\\hline
\end{tabular} \\
$^{a}$ The units are 10$^{31}$ W. \\
$^{b}$ The units are 10$^{37}$ W.\\
$^{c}$ This is the 0.5$-$2~keV unabsorbed luminosity.\\
$^{d}$ This is the 2$-$10 keV unabsorbed luminosity. \\
$^{e}$ This value is not given as the broken power-law model did not yield a good fit.
\end{table}
\end{center}

We predict the X-ray luminosity of these PG~quasars at 0.5, 2 and
5~keV (see Table~\ref{tab:tab2}). These energies were chosen so as to
sample part of the soft X-ray emission that is heavily affected by
absorption, as well as part of the hard X-ray emission, that is much
less affected. Any serious over-prediction of the 5~keV luminosity
would indicate that the absorber is close to or Compton-thick. In the
soft X-ray band, part of the emission might come directly from the
accretion disc, the so-called soft excess, and thus the difference
between predicted and measured X-ray luminosity is a lower limit,
barring X-ray variability that is larger than the soft excess
emission. Therefore, in modelling the absorption necessary to equate
the predicted and measured X-ray luminosity we will focus on the 2 and
5~keV data. The hard X-ray band is well modelled by a simple power-law
model. This power-law component is generally believed to be due to
Compton scattering of accretion disc photons. The contribution from a
reflection component should be small to negligible at 2 and 5~keV. At
higher energies, the reflection component becomes larger, therefore we
will not compare the predicted and measured 10~keV luminosity.

We note that the radio and X-ray luminosities were determined with a
difference in time of about 15 years. The X-ray luminosity is known to
be variable and this will cause a scatter in the relationship between
predicted and measured X-ray luminosity. The X-ray luminosity
variability of Seyfert 1 galaxies depends on the mass of the central
black hole, with lower mass objects generally being more variable, and
on shorter timescales \citep{uttley02,mchardy04}. For the PG~quasars
studied here, the black hole masses have been determined with a
variety of methods and range between 2$\times$10$^{7}$ M$_{\odot}$
(PG~1440+356) and 2$\times$10$^9$ M$_{\odot}$ (PG~1425+267)
\citep{Brocksopp06}, and on average are larger than the well studied
Seyfert~1 galaxies. Seyfert~1 galaxies are known to have an ``off''
state, where the luminosity is about an order of a magnitude
lower. However, these off-states are rather rare occurrences, and we
would not expect more than 2 of the studied PG~quasars to be in such a
state. Therefore, luminosity variability cannot explain a systematic
over-prediction of the X-ray luminosity.

To compare to the observed luminosities, which reported as
$\nu$L$_{\nu}$, we afterwards multiplied the predicted luminosity by
the corresponding frequency of the emission. For the rest of the paper
the luminosities quoted are $\nu$L$_{\nu}$. \cite{Brocksopp06}
modelled the 2$-$5~keV spectrum with a simple power-law model with
Galactic absorption, which gave a good fit to this part of the
spectrum. However, this fit was inadequate to fit the whole
0.3$-$10~keV spectrum. To fit the whole energy range they used a
broken power-law model with Galactic absorption, listing in their
table 3, the 2 photon indices, the break energy, the 0.3$-$10~keV
luminosity and the column density, or upper limit, for a neutral
absorber presumed part of the quasar.

Due to the listed model parameters we can calculate the observed 0.5
(0.4$-$0.6), 2 (1.9$-$2.1) and 5 (4.9$-$5.1)~keV luminosity, which can
be directly compared to the 0.5, 2 and 5~keV predicted
luminosities. Because the information about the spectral model of the
X-ray data is not given by \cite{LaorBehar08} or the references they
refer to, this was not possible for the complete set studied by
them. Considering the range of power-law photon indices determined by
\cite{Brocksopp06}, between 0.34 and 2.54, correcting the 0.2$-$1~keV
luminosity (Behar, priv. comm. 2010) given by \cite{LaorBehar08} to
either the 0.5 or 2~keV luminosity will result in ambiguous results,
and is the main reason we use a smaller sample in this study.

To be specific, we used the 2$-$5~keV luminosity with the simple
power-law model to calculate the 2 and 5~keV luminosity using {\tt
  SPEX}\footnote{http://www.sron.nl/spex}. To calculate the 0.5~keV
$\nu$L$_{\nu}$ we used the broken power-law model and the 0.3$-$10~keV
$\nu$L$_{\nu}$ given by \cite{Brocksopp06}. For PG~0050+124,
PG~0844+349 and PG~2214+139 we use the continuum model and parameters
provided by \cite{Costantini08, gallo10,
  Piconcelli04}. Table~\ref{tab:tab2} lists the predicted X-ray
luminosities as well as the 0.5, 2 and 5~keV luminosity calculated
from the spectral models and parameters. For these 3 energies the
predicted luminosity is larger than the measured one, over-predicting
the measured luminosity by factors ranging from 2.8 and
2.2$\times$10$^4$. Table~\ref{tab:tab2} lists the differences between
the predicted and measured 0.5, 2 and 5~keV luminosity in log values.

\begin{center}
\begin{figure}
\includegraphics[width=5truecm,angle=-90]{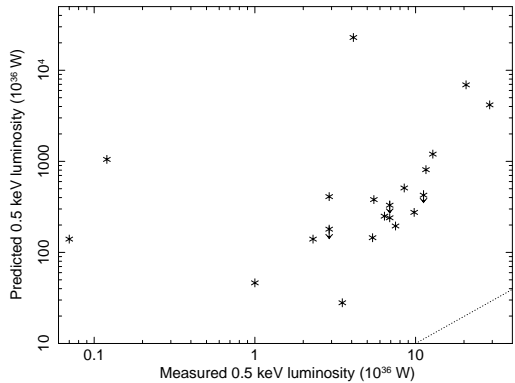}
\caption{Predicted bremsstrahlung luminosity and measured luminosity
  at $0.5$~keV for the PG quasars studied. The dotted line indicates
  where these values are equal.}
\label{fig:diff1}
\end{figure}
\end{center}

\begin{center}
\begin{figure}
\includegraphics[width=5truecm,angle=-90]{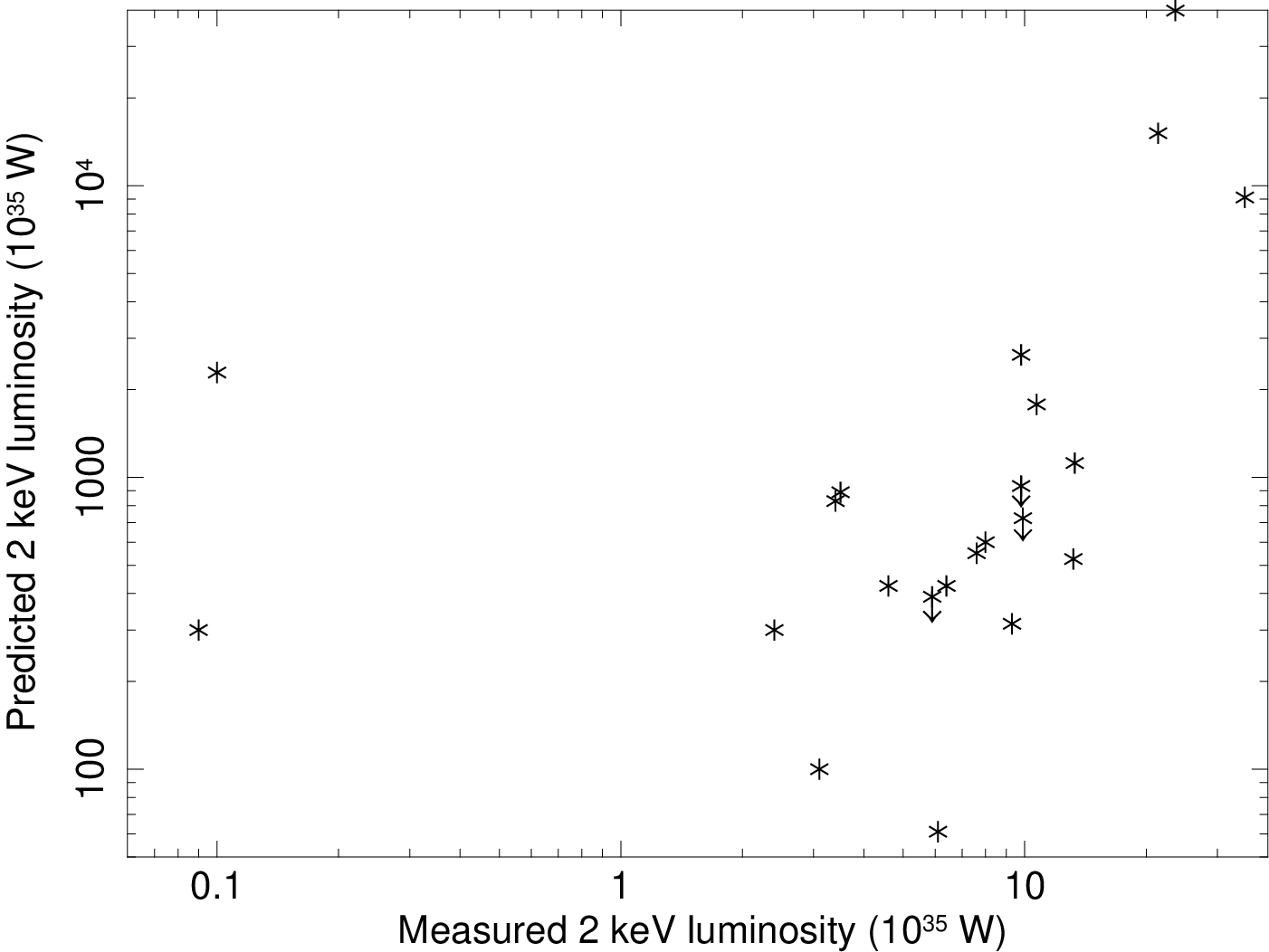}
\caption{Same as Fig.~\ref{fig:diff1}, but for $2$~keV.}
\label{fig:diff2}
\end{figure}
\end{center}

\begin{center}
\begin{figure}
\includegraphics[width=5truecm,angle=-90]{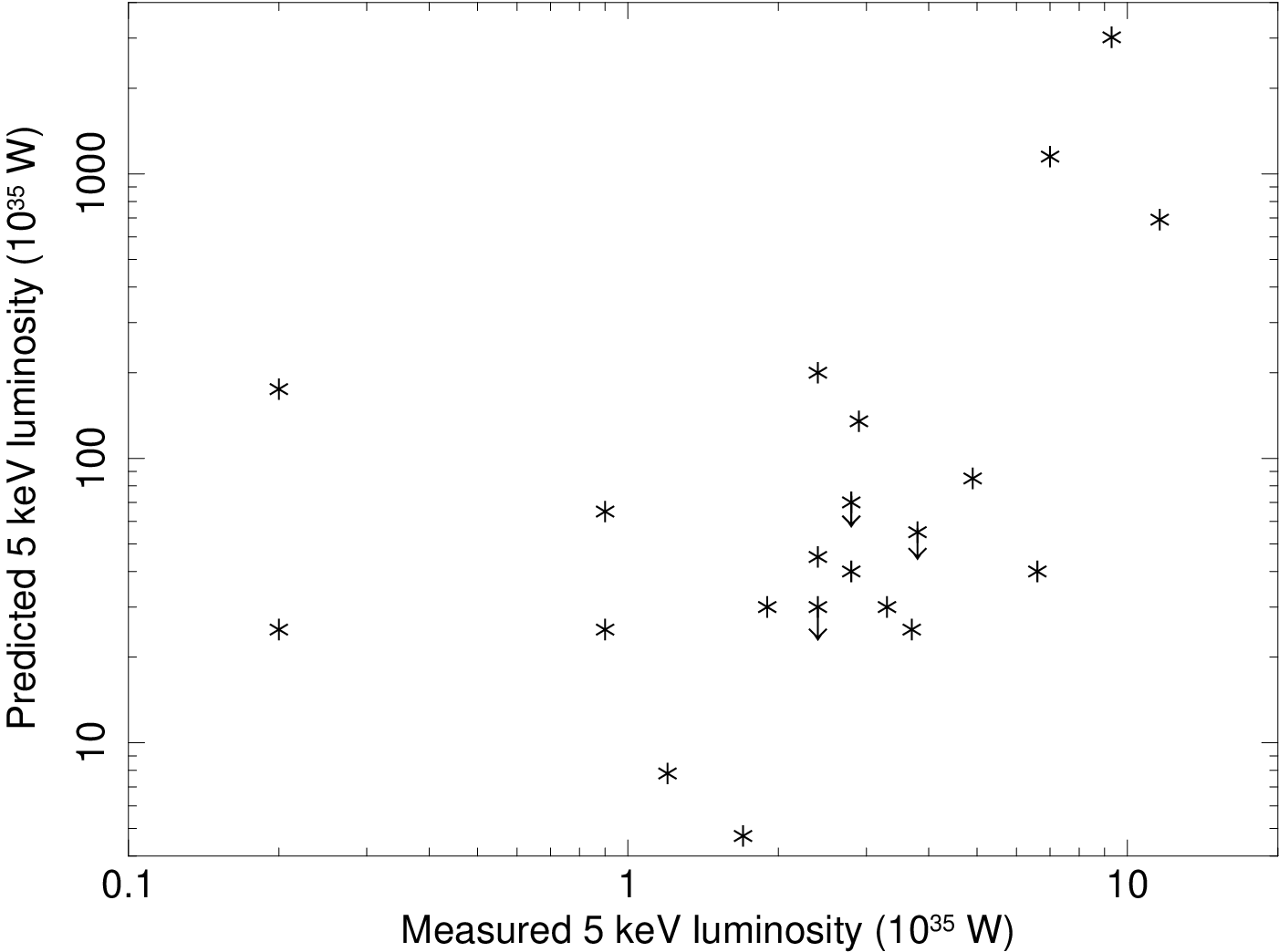}
\caption{Same as Fig.~\ref{fig:diff1} but for $5$~keV.}
\label{fig:diff3}
\end{figure}
\end{center}

For type 1 quasars, which have moderate or no absorption, if the
difference between predicted and measured $\nu$L$_{\nu}$ is due to
absorption, then one expects that the difference to be largest for
0.5~keV and smallest for 5~keV. For the moderate column densities or
upper limits listed by \cite{Brocksopp06}, one would expect that these
PG quasars are indeed all type 1 (but see later for the 2 possible
exceptions). Absorption, both neutral and ionised, affects mostly the
soft X-ray band and if moderate should become negligible above
2~keV. This is clearly not observed from Table~\ref{tab:tab2}: the
difference between predicted and measured $\nu$L$_{\nu}$ (in log) at
2~keV is larger than for 0.5~keV, and thus certainly not negligible. A
likely explanation for why the luminosity difference is larger, is the
soft excess emission which we sample at 0.5~keV but not at 2~keV. On
average in both bands the difference between predicted and measured
luminosity is more than 2 orders of magnitude. The difference between
predicted and measured $\nu$L$_{\nu}$ at 5~keV, which is least
affected by absorption, is smaller but still
significant. Figs.~\ref{fig:diff1}, \ref{fig:diff2} and
\ref{fig:diff3} show the measured versus the predicted X-ray
luminosity for 0.5, 2 and 5~keV.

\begin{center}
\begin{table*}
\caption{List of measured and predicted X-ray monochromatic
  luminosity, $\nu$L$_\nu$ using eq.~2 at 0.5, 2 and 5~keV. Also
  listed is the difference between predicted and measured X-ray
  luminosity in log, as well as the difference in measured luminosity
  for the neutral absorber column density quoted by Brocksopp et
  al. (2006). The last four columns list the hydrogen column density
  of a neutral and ionised (log$\xi$ = 1.5) absorber needed to equate
  the measured and predicted 2 and 5~keV luminosity. For PG~0050+124,
  PG~1001+054, PG~1114+445,PG~1411+442 and PG~2214+139, we used the
  detailed warm absorber models to model the neutral and ionised
  column density listed in the last 2 columns. }
\label{tab:tab2}
\begin{tabular}{c|ccc|ccc|ccc|ccccc}\hline\hline
 & \multicolumn{3}{c}{calculated}
 & \multicolumn{3}{c}{predicted} 
 & \multicolumn{3}{c}{difference}  \\
name        & $\nu$L$_{0.5}^{a}$ & $\nu$L$_2^{b}$ & $\nu$L$_5^{b}$ & $\nu$L$_{0.5}^{a}$ & $\nu$L$_2^{b}$ & $\nu$L$_5^{b}$ & $\nu$L$_{0.5}$ & $\nu$L$_{2}$ & $\nu$L$_{5}$ & abs. $\nu$L$_5^{c}$ & N$_{\rm H,n,2}^{d}$ & N$_{\rm H,i,2}^{d}$ & N$_{\rm H,n,5}^{d}$ & N$_{\rm H,i,5}^{d}$\\\hline
PG~0050+124& 2.9 & 3.5 & $-$ & 410   & 890   & $-$  & 2.1  & 2.4  & $-$  & $-$ & 0.13 & 0.26 &$-$ & $-$\\
PG~0844+349& 3.5 & 6.1 & 1.7 & 28    & 61    & 4.7  & 0.9  & 1.0  & 0.44 & $-$ & 0.05 & 0.11 & 0.28 & 0.32 \\ 
PG~0947+396& 8.5 & 13.3& 4.9 & 510   & 1120  & 85   & 1.8  & 1.9  & 1.2  &$<$0.9& 0.10 & 0.20 & 0.74 & 0.9\\
PG~0953+414& 28.9& 35.1& 11.6& 4170  & 9120  & 690  & 2.2  & 2.4  & 1.8  &$<$2.3& 0.13 & 0.26 & 1.15 & 1.3\\
PG~1001+054& 0.12& 0.1 & 0.2 & 1050  & 2290  & 175  & 3.9  & 4.3  & 2.9  &$<$4.8 & 0.2 & 0.41 &1.3 & 1.3\\
PG~1048+342& 2.9 & 5.9 & 2.4 & $<$180& $<$390& $<$30&$<$1.8&$<$1.8&$<$1.1&$<$2.5& $<$0.1 & $<$0.2 & $<$0.7 & $<$0.8\\
PG~1114+445& $-$ & 4.6 & 3.3 & 195   & 425   & 30   & $-$  & 2.0  & 1.0  & $-$ &0.09 & 0.19 & 0.25 & 0.75\\
PG~1115+407& 7.5 & 6.4 & 1.9 & 195   & 425   & 30   & 1.4  & 1.8  & 1.2  & $<$4.4 & 0.10 & 0.19 & 0.75 & 0.85\\
PG~1116+215& 20.6& 21.4& 7.0 & 6920  & 15135 & 1150 & 2.5  & 2.8  & 2.2  & $<$2.6 & 0.15 & 0.31 & 1.45 & 1.6\\
PG~1202+281& 6.9 & 13.2& 6.6 & 240   & 525   & 40   & 1.5  & 1.6  & 0.8  & $<$0.8 & 0.09 & 0.17 & 0.5 & 0.6\\
PG~1216+069& 4.1 & 23.6& 9.3 & 22910 & 39810 & 3020 & 3.7  & 3.2  & 2.5  & 6.5 & 0.17 & 0.35 & 1.65 & 1.8\\
PG~1322+659& 9.8 & 8.0 & 2.4 & 275   & 600   & 45   & 1.4  & 1.9  & 1.3  & $<$4.4 & 0.10 & 0.21 & 0.85 & 0.95\\
PG~1352+183& 6.4 & 7.6 & 2.8 & 250   & 550   & 40   & 1.6  & 1.9  & 1.2  & $<$0.8 & 0.1 & 0.21 & 0.8 & 0.85\\
PG~1402+261& 11.6& 10.7& 2.9 & 810   & 1780  & 135  & 1.8  & 2.2  & 1.7  & $<$2.0 & 0.12 & 0.24 & 1.1 & 1.25 \\
PG~1411+442& 0.07& 0.09& 0.2 & 140   & 300   & 25   & 3.2  & 3.5  & 2.1  & $<$0.03 & 0 & 0 & 1.77 & 1.5 \\
PG~1415+451& 2.3 & 2.4 & 0.9 & 140   & 300   & 25   & 1.8  & 2.1  & 1.4  & $<$1.4 & 0.11 & 0.23 & 0.9 & 1\\
PG~1427+480& 6.9 & 9.9 & 3.8 & $<$330& $<$725& $<$55&$<$1.7&$<$1.9&$<$1.2& $<$1.9 & $<$0.10 & $<$0.21 & $<$0.8 & $<$0.9 \\
PG~1440+356& 5.5 & 3.4 & 0.9 & 380   & 830   & 65   & 1.8  & 2.4  & 1.8  & 1.8 & 0.13 & 0.26 & 1.2 & 1.3\\
PG~1444+407& 11.2& 9.8 & 2.8 & $<$425& $<$935& $<$70&$<$1.6&$<$2.0&$<$1.4&$<$2.2 & $<$0.11 & $<$0.22  & $<$0.9 & $<$1\\
PG~1543+489& 12.8& 9.8 & 2.4 & 1200  & 2630  & 200  & 2.0  & 2.4  & 1.9  & $<$49 & 0.13 & 0.26 & 1.25 & 1.4\\
PG~1626+554& 5.4 & 9.3 & 3.7 & 145   & 315   & 25   & 1.4  & 1.5  & 0.8  & $<$3.3 & 0.08 & 0.16 & 0.5 & 0.6\\
PG~2214+139& 1.0 & 3.1 & 1.2 & 46.3  & 100   & 7.8  & 1.7  & 1.5  & 0.8  & $-$   & 0.08 & 0.17 & 0.5 & 0.6 \\\hline
\end{tabular} \\
$^{a}$ The units are 10$^{36}$ W.\\
$^{b}$ The units are 10$^{35}$ W.\\
$^{c}$ This is in 10$^{-4}$. \\
$^{d}$ The units are 10$^{28}$ m$^{-2}$. \\
\end{table*}
\end{center}

\section{X-ray Absorption}\label{abs}

The optical depth needed for the predicted X-ray luminosity $L_\nu$ to
match that observed $L_{\nu, {\rm obs}}$ is
\begin{equation}
 \tau = N_{\rm H} \sigma = \ln \left( \frac{L_\nu}{L_{\nu{\rm, obs}}} \right)
\end{equation}
where $N_{\rm H}$ is the absorbing hydrogen column density and
$\sigma$ is the atomic cross section. The ionisation parameter is
\citep{Tarter69}
\begin{equation}
 \xi = \frac{L}{n_{\rm H}r^2}
\end{equation}
where $L$ is the 1$-$1000 Rydberg luminosity, $n_{\rm H}$ is the
hydrogen density of the illuminated gas, and $r$ is the distance of
the absorber from the ionising source, presumed to be the inner
accretion disc. The ionisation parameter in the well studied Seyfert~1
galaxies covers a large range: $0.1 < \xi < 10^4$, with multiple
ionisation parameters needed to adequately model the absorber
\citep{Steenbrugge03,Steenbrugge05,Costantini07,Holczer07,Behar09}.

To test whether absorption is causing the difference between the
predicted and measured X-ray luminosity, we modelled with {\tt SPEX}
the luminosity difference at 5~keV caused by a neutral absorber which
has a column density listed in table 3 of \cite{Brocksopp06}. For the
neutral absorber we used the {\it abs} model, while we used the {\it
  xabs} model for an ionised absorber. The {\it abs} model uses the
\cite{mccammon90} cross-sections for a neutral absorber. {\it xabs}
calculates the continuum and line absorption in a self-consistent
fashion for all the ions for a given ionisation parameters. The depth
of the absorption is then scaled with the hydrogen column density. We
assumed solar abundances given by \cite{AndersGrevesse89}.

For those PG quasars where only an upper limit to the column density
is listed, we used this upper limit. We used the photon index given
for the simple power-law model in our modelling, similar to how we
determined the 5~keV measured luminosity. The resulting difference in
luminosities is, as expected, very small, on average (in log) 0.0005,
and much smaller than the difference between predicted and measured
luminosity at 5~keV. As for most PG quasars \cite{Brocksopp06} gives
only upper limits, the real difference in luminosity due to absorption
should be even smaller. Therefore, either the model, i.e. the fact
that the radio and X-ray emission arise from a wind, is wrong, or the
absorption in these systems is severely underestimated by
\cite{Brocksopp06}. The latter is possible if the broken power-law
model, which fits the soft excess also fits the absorption. For
absorbed sources one expects that the soft X-ray photon index is
flatter than the hard X-ray photon index due to absorption. This is
not the case, but this could be due to a differing shape of the soft
excess in these sources compared to the soft excess in Seyfert~1
galaxies.

There is one PG quasar that has a high-resolution reflection grating
spectrometer (RGS) spectrum available and which is listed by Laor and
Behar, PG~0050+124 (also known as IZwicky 1), we therefore include it
in our sample. The spectrum is well modelled by a 2 component warm
absorber with log($\xi$/10$^{-9}$ Wm) = 0 and 2.6 and a N$_{\rm H}$ =
13.3$\times$10$^{24}$ and 13.5$\times$10$^{24}$ m$^{-2}$, respectively
\citep{Costantini08}. These authors also list the unabsorbed
(i.e. correcting the measured luminosity for the absorption
components) fitted between 0.5$-$2 keV luminosity as well as the
power-law slope using the RGS data. Therefore, we can calculate the
0.5 and 2~keV luminosity and compare it to the predicted luminosity at
these energies. The difference is more than 2 orders of magnitude: 140
and 250 for the 0.5 and 2~keV luminosity respectively. As the measured
luminosity is corrected for the Galactic and intrinsic absorption
measured in this spectrum, this difference cannot be explained by the
neutral and ionised absorption in this source. Furthermore, as this is
a high-resolution spectrum, excess absorption of any significance
which is needed to reduce the difference between predicted and
measured $\nu$L$_X$ is excluded. Therefore, for this source absorption
cannot explain the difference between the predicted and measured X-ray
luminosity.

For PG~1001+054, PG~1114+445, PG~1411+442 listed by \cite{Brocksopp06}
and PG~2214+139 (also known as Mrk~304) there are more detailed
studies available using the EPIC data. For PG~1411+442 only a neutral
absorber is fitted, while for PG~1001+054, PG~1114+445 and PG~2214+139
a warm absorber model is fitted. \cite{schartel05} studied PG~1001+054
and derived a hydrogen column density of 19.2$\times$10$^{26}$
m$^{-2}$ and log($\xi$/10$^{-9}$ Wm) = 2.7. \cite{Ashton04} studied
PG~1114+445 in detail, fitting a 2 component warm absorber and noting
that the absorption parameters derived are very similar to the ones
derived for NGC~3783 using a 2 component model. They measure 2
ionisation parameters: log($\xi$/10$^{-9}$ Wm)) = 0.83 and 2.57 and
N$_{\rm H}$ = 7.41$\times$10$^{25}$ and 5.25$\times$10$^{26}$
m$^{-2}$, respectively. \cite{Piconcelli05} found the same parameters
as \cite{schartel05} for PG~1001+054 and fitted a simpler model to
PG~1114+445, which we will not discuss here. In addition,
\cite{Piconcelli05} did study 1411+442, fitting the continuum spectrum
with a neutral absorber with hydrogen column density of
2.3$\times$10$^{26}$ m$^{-2}$. \cite{Piconcelli04} fitted a two
component warm absorber model to PG~2214+139, measuring hydrogen
column densities of 17$\times$10$^{25}$ and
89$\times$10$^{25}$~m$^{-2}$ and ionisation parameters of 0.77 and
1.95. As is the case for PG~0050+124, \cite{Piconcelli04} gives the
unabsorbed X-ray luminosity for PG~2214+139. Finally, for PG~0844+349
EPIC spectra were studied in detail by \cite{brinkmann06} and
\cite{gallo10}. \cite{brinkmann06} concludes that there is possibly a
detection of the Fe~XXVI~Ly$\alpha$ absorption line, but that the
mass-loss rate is certainly less than
1~M$_{\odot}$yr$^{-1}$. \cite{gallo10} fits the different flux state
spectra assuming the same continuum model holds and prefers a model
with a power-law and blurred reflection component, where the
normalisation and photon index of the power-law component is
variable. In this model there is no absorption, but they cannot rule
out a model with a variable absorber and no reflection component. The
details of these warm absorber models, as well as for PG~0050+124, are
listed in Table~\ref{tab:tab3}.

We fitted the absorber of these 5 PG~quasars and determined the
difference in 5 keV luminosity due to the absorber. In four cases the
difference was minimal, a factor of 1.3, 1.1, 1.06, 1.2 and 2.25 for
PG~0050+124, PG~1001+054, PG~1114+445, PG~1411+442 and PG~2214+139
respectively. The difference between predicted and measured 5~keV
luminosity is a factor of 760, 9.7, 135 and 6.3, and is much larger
than is explained by the fitted absorber. To test if further
absorption could explain the luminosity difference, we modelled the
hydrogen column density of an absorber required to have an unabsorbed
5~keV luminosity equal to that predicted. Note that due to the
possible X-ray luminosity difference between when the radio data and
the X-ray data were obtained, this hydrogen column density is not an
exact number. However, it should give a conclusive result as to
whether excess absorption is a possible explanation.

For all five sources we decided to add a neutral as well as ionised
absorber to the fitted absorber model, as absorption due to a neutral
gas has a higher optical depth, and therefore requires a less large
hydrogen column density for the same amount of absorption. However, an
ionised absorber is more likely if it comes from the wind. For
PG~1114+445 the necessary column density is 0.25$\times$10$^{28}$
m$^{-2}$ for a neutral absorber and 0.6$\times$10$^{28}$ m$^{-2}$ for
an ionised absorber with an ionisation parameter of 2.57, i.e. the
same as the highly ionised component fitted by \cite{Ashton04}. Both
these column densities are larger than the best fit values derived by
\cite{Ashton04} of 7.41$\times$10$^{25}$ and 5.25$\times$10$^{26}$
m$^{-2}$. Note that the neutral and ionised hydrogen column density we
derive is in addition to the absorber fitted by
\cite{Ashton04}. Therefore, excess absorption unlikely explains the
difference between the measured and predicted luminosity in the case
of PG~1114+445. For PG~2214+139 the extra neutral hydrogen column
density is 0.6$\times$10$^{28}$ m$^{-2}$, larger than the total
hydrogen column density of 0.1$\times$10$^{28}$ m$^{-2}$
\cite{Piconcelli04} derives. Assuming an ionised absorber with
log$\xi$ = 1.95 (the high ionisation parameter) we find an extra
hydrogen column density of 0.65$\times$10$^{28}$ m$^{-2}$. Again, this
indicates that excess absorption is unlikely to explain the difference
between predicted and measured 5~keV X-ray luminosity.

For PG~1001+054 the additional neutral hydrogen column density needed
is 1.3$\times$10$^{28}$ m$^{-2}$, while for an ionised absorber with
an ionisation parameter of 2.7 \citep{schartel05} is
2.7$\times$10$^{28}$ m$^{-2}$. Even the hydrogen column density of the
neutral absorber is high enough that this would indicate a marginally
Compton-thick absorber. For PG~1411+442 we calculated the needed
neutral hydrogen column density for the unabsorbed measured 5 keV
luminosity to be 1.77$\times$10$^{28}$ m$^{-2}$, which is nearly 2
orders of magnitude more than measured and would make this also a
marginally Compton-thick spectrum. As no ionised absorber was fit for
this source by \cite{Piconcelli05}, we fitted an ionised absorber with
an canonical ionisation parameter of 1.5 (log, 10$^{-9}$ Wm), as we
also did for the remaining PG~quasars in our sample without detailed
absorber modelling.

\begin{center}
\begin{table*}
\caption{The more detailed absorption model parameters: hydrogen
  column density and ionisation parameter, and the resulting
  difference in $\nu$L$_X$ at 5~keV due to the intervening
  absorption. The next to last column lists the extra neutral hydrogen
  column density that is required to bring the predicted X-ray
  luminosity in agreement with the measured X-ray luminosity at
  5~keV.}
\label{tab:tab3}
\begin{tabular}{cccccccl}\hline\hline
name        & N$_{\rm H}^{a}$ & $\xi^{b}$ & N$_{\rm H}^{a}$ & $\xi$ & $\Delta \nu$L$_X$  & N$_{\rm H,5}^{a}$ & reference \\
PG~0050+124 & 1.33  & 0    & 1.35    & 2.6   & 0    & 130$^{c}$ & Costantini et al. 2008 \\
PG~1001+054 & 192   & 2.7  & $-$     & $-$   & 0.11 & 1300 & Schartel et al. 2005 \\
PG~1114+445 & 7.41  & 0.83 & 52.5    & 2.57  & 0.05 & 250 & Ashton et al. 2004 \\
PG~1411+442 & 23    & n$^{d}$ & $-$  & $-$   & 0.03 & 1770 & Piconcelli et al. 2005 \\ 
PG~2214+139 & 17    & 0.77 & 89     & 1.95  & 0     & 500 & Piconcelli et al. 2004 \\\hline
\end{tabular} \\
$^{a}$ In 10$^{25}$ m$^{-2}$. \\
$^{b}$ Log, and in 10$^{-9}$ W m.\\
$^{c}$ Determined using the 2~keV luminosity, as for the RGS only the 0.5$-$2~keV luminosity is given. \\
$^{d}$ The fitted absorber was assumed neutral.
\end{table*}
\end{center}

\begin{figure}
\caption{The measured neutral hydrogen absorption, or upper limit as
  listed by Brocksopp et al. (2006) versus the required neutral
  hydrogen column density for the predicted and measured 5~keV
  luminosity to be equal. We show all the PG~quasars studied here,
  with the exception of PG~0050+124, PG~1001+054, PG~1114+445,
  PG~1411+442 and PG~2214+139.}
\includegraphics[width=5truecm,angle=-90]{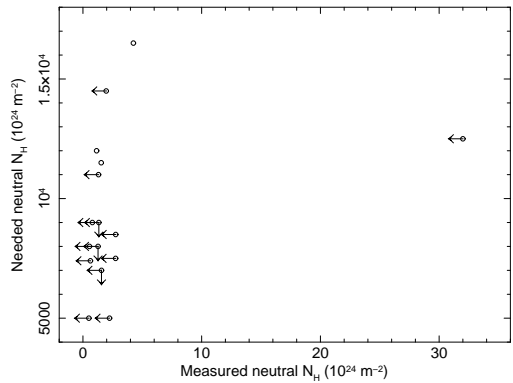}
\label{fig:abs}
\end{figure}

For all the PG~quasars in our sample, we list in Table~\ref{tab:tab2}
the difference in 5~keV luminosity due to the neutral hydrogen column
density derived by \cite{Brocksopp06}. \cite{Brocksopp06} finds mostly
upper limits to the neutral hydrogen column density due to an
intrinsic absorber, and thus the difference in luminosity is mostly
given by upper limits. In the same Table we list the hydrogen column
density a neutral (ionised) absorber needs to have to equate the
predicted and measured 2 and 5~keV luminosity. We chose to model the
neutral hydrogen column density, as the ionisation parameter of the
possible absorber is unknown, and because neutral gas is more
efficient in absorbing X-ray radiation, thus this gives the minimum
absorbing column density needed. For the ionised absorber we chose an
ionisation parameter of 1.5 (log, 10$^{-9}$ Wm), which is a rough
average of the ionisation parameters observed in the well studied
Seyfert~1 galaxies. The units for these modelled hydrogen column
densities are in 10$^{28}$ m$^{-2}$, where a column density higher
than 1.5$\times$10$^{28}$ m$^{-2}$ indicates a Compton-thick absorber
\citep{matt99}. In modelling the source we used the 2$-$5~keV
luminosity and power-law slope as given by \cite{Brocksopp06} or used
the detailed models available for the 4 sources discussed
earlier. Fig.~\ref{fig:abs} shows the required versus measured column
density of an X-ray absorber using the 5~keV luminosity. This figure
does not include the 4 quasars with detailed absrober modelling.

A recent study of nearby Seyfert~1 galaxies suggests that an absorber
produced by a large scale, continuous radial flow from an accretion
disc may have a density profile $n \propto r^{-\alpha}$ where $1 <
\alpha < 1.3$ \citep{Behar09}. In particular, for the Seyfert galaxy
NGC 5548 \citet{Steenbrugge05} find that the observed X-ray spectrum
can be described by a model with a continuous nonuniform distribution
of density and ionisation parameter, suggesting that the warm absorber
is not made up of clouds of ionised material in pressure equilibrium
with the surrounding wind. Thus our absorption model assuming just a
neutral or 1 ionisation parameter is overly simplified. In Seyfert~1
galaxies most of the gas is highly ionised \citep{Steenbrugge03},
which is less efficient in absorbing the continuum. Therefore, even
for the ionised absorber modelled here, the hydrogen column densities
derived are lower limits, if the overall ionisation structure of the
absorber in RQQs is similar to that in Seyfert~1 galaxies.

\section{Discussion}\label{discussion}

If the radio spectrum of radio-quiet quasars is produced in a
free-free disc wind, the resultant bremsstrahlung X-ray emission must
be absorbed because the observed X-ray spectrum is typically a power
law, and not an exponential, and because the predicted X-ray
luminosities for 0.5, 2 and 5~keV are much larger than those measured
for all the PG~quasars in our sample.

Studying Table~\ref{tab:tab2}, we find that the hydrogen column
density needed to absorb the 2 and 5~keV luminosity are rather
different. For the neutral (ionised) absorber the hydrogen column
density is $\sim$8 (4) times larger for the 5~keV luminosity than for
the 2~keV luminosity, although the difference between predicted and
measured luminosity is smaller at 5~keV. This difference could be
explained if a more complicated absorber is present with most of the
gas highly (i.e. $>$3 but $<$4 in log, 10$^{-9}$ Wm) ionised. However,
for such a high ionisation parameter, the required hydrogen column
density required is further increased and becomes of order
2$\times$10$^{28}$ m$^{-2}$.

We find that most of the PG~quasars would need to have an absorber
that is nearly Compton-thick to explain the difference in luminosity
between the predicted and measured 5~keV luminosity. Furthermore, to
be able to fit the 2 and 5~keV luminosity a highly ionised absorber
with very large column density is required. If the absorbers in RQQs
have similar properties to those in the well studied Seyfert~1
galaxies, which one would expect to be the case considering their very
similar X-ray properties such as the power-law slope and the
temperature of the soft excess (see for instance
\citealt{Piconcelli05}), is very unlikely. Instead, from the derived
hydrogen column densities we predict that PG~quasar spectra should be
very similar to Seyfert~2 galaxies, which show in high-resolution
X-ray spectra an emission line spectrum in the soft X-ray
band. Furthermore, they have a much flatter hard X-ray power-law
photon index if the absorption is not fit or severely
underestimated. From a statistical point of view it is unlikely that
the PG~quasar sample would be entirely consisting of those quasars
with an obscuring torus along the line of sight, even-though the used
sample is unlikely free from biases. Is there other evidence that this
is not the case? For PG~0050+124, where \cite{Costantini08} present a
high-resolution X-ray spectrum, it can be ruled out that the source is
similar to a Seyfert~2 galaxy, instead it shows a spectrum that is
very similar to a weakly absorbed Seyfert~1 galaxy. This is the only
quasar for which we have a high-resolution spectrum and for which we
can exclude that extra absorption is the cause of the difference
between the predicted and measured X-ray luminosity. The predicted
2~keV luminosity of this PG~quasar is 250 times the measured
luminosity, which is unlikely explained by luminosity
variability. Therefore, for PG~0050+124 we are certain that the radio
emission cannot be due to optically thin emission from a disc wind.

For the other quasars in our sample, we have to use a less direct
method, and that is to compare the measured photon index and
equivalent width of the Fe~K$\alpha$ emission line to those measured
in Seyfert~1 galaxies and those in Seyfert~2 and Compton-thick
AGN. The photon index for all but 2 of the studied PG~quasars is not
flat as would be expected if they had a near Compton-thick absorber,
but varies around the photon indices measured in Seyfert 1 galaxies,
i.e. $\Gamma$ = 1.9 ($<\Gamma>$ = 2.1 for the fitted quasars minus
PG~1001+054 and PG~1411+442 see table~3 in \citealt{Brocksopp06}). The
equivalent width of the Fe~K$\alpha$ line stated by \cite{Brocksopp06}
is rather poorly determined and dependent on the assumed width of the
line, but is generally, with the exception of PG~1001+054 and
PG~1411+442, consistent with being too small for a Compton-thick
absorption model. Finally, we can use the UV measured equivalent width
of the C~IV 1549~\AA~absorption line. In our sample only 4 quasars
show an C~IV equivalent width that is larger than 1~\AA, although a
measurement was made in all but one of the 22 quasars. As there is
generally a good correlation between UV and X-ray measured absorption,
this would indicate that only 4 quasars in our sample have significant
X-ray absorption. From the combination of the above arguments we
conclude that for all but 2 of the studied PG~quasars (PG~1001+054 and
PG~1411+442) an absorber with a very large hydrogen column density can
be ruled out as an explanation for the difference between predicted
and measured X-ray luminosity.

Could PG~1001+054 and PG~1411+442, the 2 PG~quasars with a flat hard
X-ray power-law as measured by \cite{Brocksopp06} have a nearly
Compton-thick absorber? Signatures in medium-resolution spectra of
Compton-thick spectra are a large equivalent width of the Fe~K$\alpha$
line and a flat hard energy spectrum, which in our case is the 2$-$5
keV spectrum. Indeed for these 2 PG~quasars \cite{Brocksopp06} finds a
flat hard energy photon index: 0.14$\pm$0.4 and 0.35$\pm$0.1,
respectively. For the broken power-law model fitted between
0.3$-$10~keV, the high energy photon indices for these 2 RQQs is 0.74
and 0.34, and thus still flat. Consistent with an origin of the flat
photon indices due to absorption \cite{schartel05} and
\cite{Piconcelli05} do not derive such a flat power-law indices in
their best fit model with absorber, instead they derive a photon index
very similar to the canonical 1.9 measured in Seyfert~1 galaxies: 1.97
and 1.9 for PG~1001+054 and PG~1411+442, respectively.

A second signature is the equivalent width of the Fe~K$\alpha$
emission line, which ranges between $<$2 and $<$ 10$^6$ and $<$428 and
$<$706 eV \citep{Brocksopp06}, and thus could be large, but is too
poorly determined to be a constraint. \cite{schartel05} note that the
spectrum of PG~1001+054 has a low signal-to-noise, and that the fitted
model is a poor fit ($\chi$ = 15.6) to the data, but that a more
complicated model is not warranted considering that only 12 energy
bins remain after data binning. Therefore, no Fe~K$\alpha$ line was
fitted. \cite{Piconcelli05} does fit a narrow Fe~K$\alpha$ emission
line and find that the emission is neutral, but does not mention the
measured equivalent width of this line. However, as for PG~1001+054,
this is a rather low signal-to-noise spectrum, and \cite{Piconcelli05}
claims that even fitting an ionised absorber is not warranted by the
data quality.

PG~1001+054 is classified as a narrow line quasar based on the
full-width-half-maximum (FWHM) of the H$\beta$ line of 1740~kms$^{-1}$
\citep{wills00}, indicating that the broad emission lines are more
narrow, i.e. less than 2000~kms$^{-1}$, than for most type 1 AGN. This
quasar is also classified as a broad absorption line (BAL)QSO, with
strong UV absorption, and thus likely is severely absorbed in the
X-ray band. \cite{blustin08} studied 5 X-ray selected BALQSO's and
derived hydrogen column densities between 3.2$\times$10$^{26}$
m$^{-2}$ for the one neutral absorber, and 4$\times$10$^{28}$ m$^{-2}$
for an ionised absorber. The ionised hydrogen column densities are
somewhat larger than the hydrogen column density we require for
PG~1001+054 and PG~1411+442 to match the predicted 5~keV
luminosity. PG~1411+442 shows strong UV absorption, and is with
PG~1001+054 one of the 4 sources in our sample that has a C~IV
equivalent width larger than 1~\AA. Therefore we conclude that
P~1001+054 and PG~1411+442 likely harbour a nearly Compton-thick
absorber, and that the difference between predicted and measured X-ray
luminosity could be due to absorption. For a more conclusive result
higher signal-to-noise X-ray spectra of these 2 sources need to be
obtained.

There is another high-resolution X-ray spectrum taken with the
high-energy transmission grating (HETG) on-board {\it Chandra} of the
radio-quiet quasar, MR~2251-178, which however is not listed in the
Palomar-Green catalogue. \cite{gibson05} fit the warm absorber with a
column density of 2.37$\times$10$^{25}$ m$^{-2}$ and an ionisation
parameter of 0.02 (log 10$^{-9}$ Wm). They also detect a high velocity
outflow, however, this line was not observed in the XMM-{\it Newton}
spectra \citep{kaspi04} of the source, and is thus either very
variable or a spurious detection. \cite{kaspi04} studying the
high-resolution RGS spectrum of the same source fits the absorption
with a 2 component model which have a column density of
3$-$6$\times$10$^{25}$ m$^{-2}$ for the high ionisation component and
2$\times$10$^{24}$ m$^{-2}$ for the low ionisation component. The
derived column densities for this source are higher than the average
column density derived for the PG~quasars studied by
\cite{Brocksopp06}, but similar to the column density derived by
\cite{Costantini08} for PG~0050+124 and smaller than the column
density derived by \cite{schartel05, Ashton04, Piconcelli05,
  Piconcelli04} for PG~1001+054, PG~1114+445, PG~1411+442 and
PG~2214+139, respectively.

So far we have assumed that the hard X-ray emission is directly due to
the emission from a wind with $\tau$~$\lesssim$~1. We have ignored the
fact that the observed spectrum is a power-law and not an exponential,
as predicted by Eq.~2. We have also ignored that likely at least part
of the hard X-ray emission is due to Comptionised accretion disc
photons. Indeed, if part of the hard X-ray emission does not originate
in the wind, the luminosity difference that needs to be explained is
even larger, and therefore the required absorber column densities will
be larger. This would make an absorber as an explanation between the
observed and predicted luminosities even less likely. We have further
assumed that there is no redistribution of the X-ray luminosities, as
caused by reflection. However, reflection is generally assumed to add
an insignificant luminosity at 2 and 5~keV, and should therefore not
greatly alter the column densities derived in this paper.

An added uncertainty in comparing radio and X-ray data taken years
apart is the luminosity variability observed in the X-ray part of the
spectrum. X-ray luminosity variability is random and therefore we
expect as many sources that have a smaller X-ray luminosity than that
predicted from the observed radio emission, than have a higher X-ray
luminosity. The X-ray variability probably explains part of the
difference in the derived hydrogen column densities. However, it
cannot explain the over-predicted hydrogen column density in all the
PG~quasars studied.

 


\citet{BlustinFabian09} calculated the 1.4~GHz radio emission from the
X-ray absorber measured in 5 nearby AGN. They estimate that the X-ray
absorber is optically thick, and therefore use the formalism of
\cite{wright75} to calculate the expected 1.4~GHz radio emission from
the X-ray observed absorber, assuming a spherically symmetric
wind. From the comparison between the calculated radio emission and
the observed radio emission they derive the upper limits to the volume
filling factor of the absorber. These are upper limits, because any
1.4~GHz emission from the UV part of the wind, the base of the jet,
the accretion disc or the host galaxy is ignored. The upper limits
derived range between 10$^{-4}$ and 0.5, indicating that at least is
some AGN the volume filling factor of the absorber is small. For at
least 1 AGN, NGC~3783, the radio emission cannot be explained by the
observed absorber components. It therefore seems unlikely that the
radio and X-ray emission solely come from the same wind, whether
optically thin or thick, as detected in the UV and X-ray through
absorption.


\section{Conclusions}\label{conclusion}

We have tested the theory that bremsstrahlung emission from an
optically thin disc wind can explain the radio emission in radio-quiet
quasars, by comparing the predicted X-ray emission at 0.5, 2 and 5~keV
to the measured X-ray luminosities at these energies. We have modelled
the difference in luminosity as due to a neutral or ionised absorber,
likely formed in the same disc wind, but further out. We find that all
the disc winds in the PG~quasars studied would need to have an
absorbing hydrogen column density that is nearly Compton-thick. For 20
of the 22 PG~quasars we studied we can exclude such a large hydrogen
column density from the existing medium-resolution X-ray spectra and
the small C~IV equivalent width measured in those RQQs. For one RQQ,
PG~0050+124, there is a high-resolution spectrum, which is clearly
inconsistent with the required extra absorption. For the remaining 2
RQQs: PG~1001+054 and PG~1411+442 we conclude that the spectra are consistent with the necessary absorption to explain the
difference between predicted and measured X-ray luminosity, and that
thus in these 2 quasars emission from a disc wind could explain the
radio emission from the nucleus.

\section{acknowledgments}

EJDJ acknowledges support from a University of Sydney Postgraduate
Award.  The authors would like to thank Dr Roberto Soria for useful
discussions and the anonymous referee for helpful comments.

\label{lastpage}
\end{document}